\documentclass[aps,prb,showpacs,superscriptaddress,twocolumn]{revtex4-1}

\usepackage{amsfonts}
\usepackage{subfigure}
\usepackage{amsmath}
\usepackage{txfonts}
\usepackage{amssymb}
\usepackage{amsbsy} %
\usepackage{epsfig}
\usepackage{graphicx}
\usepackage{epstopdf}

\usepackage{mathdots} %

\def\be{\begin{equation}} \def\ee{\end{equation}}
\def\bea{\begin{eqnarray}} \def\eea{\end{eqnarray}}

\def\be{{\bf e}}

\newcommand{\bra}[1]{\langle#1|}
\newcommand{\ket}[1]{|#1\rangle}

\begin{document}
\title{Dynamical quantum phase transitions on cross-stitch flat band networks}

\author{Tong Liu}
\affiliation{Department of Physics, Southeast University, Nanjing 211189, China}

\author{Hao Guo}
\email{guohao.ph@seu.edu.cn}
\affiliation{Department of Physics, Southeast University, Nanjing 211189, China}

\date{\today}

\begin{abstract}
We study the quench dynamics on cross-stitch flat band networks by a sudden change of the inter-cell hopping strength $J$. For quench processes with $J$ changing as $J=0\rightarrow J\neq0$, we give the analytical expression to the Loschmidt echo which possesses a series of zero points at critical times $t^{*}$, indicating where the dynamical quantum phase transitions occur. We further study the converse quench process with $J\neq0\rightarrow J=0$, and find a non-trivial example that the pre-quench quantum state is not an eigenstate of the post-quench Hamiltonian, whereas the Loschmidt echo $\mathcal{L}(t)\equiv1$ during this process. For both situations, these results are also illustrated numerically. Finally, we give a brief discussion on the possible experimental observation of these predictions in the system of ultracold atoms in optical lattices.
\end{abstract}


\maketitle
\section{Introduction}
\label{n1}
Phase transition, the transformation from one (equilibrium) physical state to another, is a central research topic in condensed matter physics.
The dynamical quantum phase transition (DQPT)~\cite{Heyl1,Heyl2,Heyl3}, a generalization of this
fundamental concept to the nonequilibrium quantum evolution, has been studied intensively in recent years~\cite{FA,Eck,CK,MM,Jpg,Lang}. It has been confirmed that
DQPTs are directly connected to the underlying equilibrium phase transitions of the systems in broken-symmetry phases~\cite{Fisher,Heyl,Brandner,Marzolino,Vajna1}.
For noninteracting topological systems~\cite{Vajna2,Heyl4,avb,chen1,sch,Sedlmayr}, it has been verified on general grounds that two topologically different equilibrium ground states necessarily impose the existence of DQPTs. Inspired by the underlying phase transitions in these equilibrium systems, DQPTs have also been connected to the inhomogeneous systems, including the Anderson model~\cite{wang} and the incommensurate Aubry-Andr\'{e} model~\cite{chen2}.
Recently, DQPTs have been experimentally observed in two types of quantum simulating platforms, the trapped-ion system in which the dynamics of transverse-field Ising models~\cite{Jurcevic} is synthesized, and the ultracold-atom system in which the dynamical topological quantum phase transitions are observed~\cite{Fl}.

Generically,
it is argued that the occurrence of DQPTs requires that the quench process is ramped through a quantum critical point.
For quenches not belonging to these classes, the so called ``accidental" DQPTs can still occur, requiring a fine-tuning of the Hamiltonian. In this paper we propose a quench scheme on flat band networks in which DQPTs can occur without ramping through a quantum critical point.

Flat band networks~\cite{Flach1,Flach2,Flach3,Flach4,Flach5,Gneiting} are translationally invariant tight-binding lattices which support at least one dispersionless band in the energy spectrum.
This system has usually been considered as an ideal playground to explore the strong correlation phenomena~\cite{Vidal1,Vidal2} due to the complete quenching of the kinetic energy. For example, a nearly flat
band with non-trivial topological properties was proposed to simulate fractional Chern insulators~\cite{LIU}.
For single-particle systems, Ref.~\cite{chen} argued that there exist three criteria to determine the topological properties of flat bands in two-dimensional lattices, exactly flat band, non-zero Chern number, and local hopping. The authors have demonstrated that only two criteria can be simultaneously satisfied.
The fact that all three criteria can not be satisfied simultaneously indicates that the topology of the strictly flat band of real materials (short-ranged hopping) is trivial.
Thus, the theory of DQPTs in topological band systems can not be applied to flat band systems. However, inspired by the theory of DQPTs in inhomogeneous systems~\cite{wang,chen2}, we notice a remarkable feature of the flat band,  the so-called compact localized states (CLSs)~\cite{Flach6,Flach7,Flach8,MRN,Perchikov,Vicencio1,Vicencio2}, which are strictly localized eigenstates in real space. The CLSs can be considered as a Wannier function of which the amplitude is finite only in very limited regions, and vanishes identically outside.
Contrast to the Anderson localization in which the exponentially localized states are induced by disorders, the CLSs typically occur in perfectly periodic systems, originated from the destructive interference by the hopping processes of specific lattices.

Among a wide variety of flat band networks, the classification of flat bands is useful for choosing the appropriate model to realize DQPTs. A first attempt to classify flat bands by the properties of CLSs was discussed in Ref.~\cite{Flach1}. The authors classify the CLSs by the number $U$ of unit cells occupied by a CLS. And a very recent work~\cite{Rhim} developed another classification scheme of flat band systems from the perspective of the Bloch wave function's singularity. For the $U = 1$ class, the CLSs form a set of orthogonal and complete bases~\cite{Flach1}, indicating that a single CLS is disentangled from the rest of unit cells, such as the cross-stitch network. However, for generic $U > 1$ classes, the CLSs are not orthogonal to each other anymore in one dimension, such as the sawtooth network~\cite{Huber}, and do not form a complete set of bases spanning the whole network in two dimension, such as the Lieb lattice~\cite{Lieb}, where compact localized lines must be added to form a complete set. The orthogonality and completeness of CLSs are crucial for proving the existence of zeros of Loschmidt echo in our quench protocol, which is the main motivation for choosing cross-stitch networks.

The rest of the paper is organized as follows. In Sec.~\ref{n2}, we introduce the model and study the quench dynamics with the inter-cell hopping strength $J$ changing as $J=0\rightarrow J\neq0$. We give the analytical expressions of the Loschmidt echo and demonstrate that there are a series of zeros of Loschmidt echo, which indicates the DQPTs indeed occur. In Sec.~\ref{n3}, we study the inverse quench dynamics with $J$ changing as $J\neq0\rightarrow J =0 $. We find that there exists a non-trivial example that the pre-quench quantum state is not an eigenstate of the post-quench Hamiltonian, whereas the value of Loschmidt echo is always 1 as time varies. We conclude and discuss possible experimental realizations in Sec.~\ref{n4}.

\section{MODEL AND DQPTs}
\label{n2}
\begin{figure}
  \centering
  \includegraphics[width=0.5\textwidth]{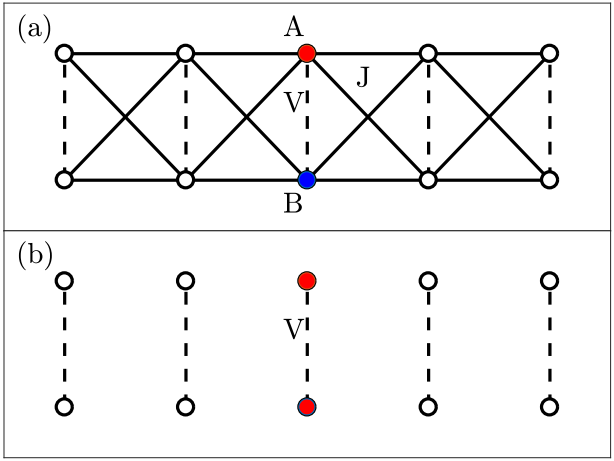}\\
  \caption{(Color online) (a) The cross-stitch geometry with the inter-cell hopping strength $J\neq0$ and the intra-cell hopping strength $V$ in the unit cell. The red and blue filled circles denote an antisymmetrical CLS $|f_{m}\rangle=(-1,1)^T \delta_{n,m} / \sqrt{2}$ localized at two sites $(A,B)$. (b) The isolated two-site geometry with the inter-cell hopping strength $J=0$ and the intra-cell hopping strength $V$. The red filled circles denote an symmetrical CLS $|\phi_-\rangle = (1,1)^T \delta_{n,m}/\sqrt{2}$.}
  \label{001}
\end{figure}

\begin{figure}
  \centering
  \includegraphics[width=0.5\textwidth]{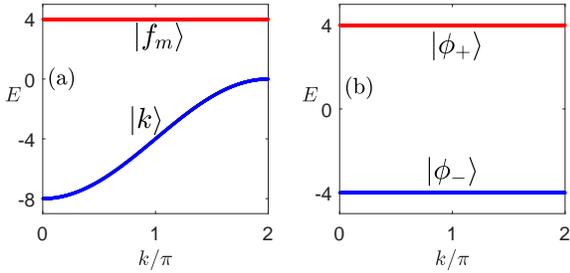}\\
  \caption{(Color online) (a) Single-particle dispersion with the corresponding eigenstates $|f_{m}\rangle$ and $|k\rangle$ on the cross-stitch lattice ($J=1$) as a function of quasimomentum $k$. (b) Single-particle dispersion with the corresponding eigenstates $|\phi_+\rangle$ and $|\phi_-\rangle$ on the isolated two-site lattice ($J=0$) as a function of the quasimomentum $k$. Here we apply the periodic boundary condition and choose $V=4$. The total number of unit cells is set to be $L = 1000$.}
  \label{002}
\end{figure}

As defined in the seminal paper~\cite{Heyl}, a key quantity within the theory of DQPTs
is the Loschmidt amplitude
\begin{equation}
 \mathcal{G}(t)= \langle \Psi_i|\Psi_i(t)\rangle= \bra{\Psi_i}e^{-i\hat H_f t} \ket{\Psi_i},
 \label{Eq1}
\end{equation}
where $\ket{\Psi_i}$ denotes the pre-quench quantum state and $\hat H_f$
the post-quench Hamiltonian. The Loschmidt echo $\mathcal{L}(t)$ is defined as the squared modulus of the Loschmidt amplitude  $\mathcal{L}(t)=|\mathcal{G}(t)|^2$.
Analogous to the equilibrium phase transition theory, the Loschmidt amplitude can be viewed as a boundary partition function along the complex temperature. And the initial state $\ket{\Psi_i}$ plays the role of a boundary condition in time instead of space. Thus, a dynamical free energy density can be defined as
\begin{equation}
 f(t) = - \lim_{L\to \infty} \frac{1}{L} \ln \mathcal{L}(t),
 \label{Eq2}
\end{equation}
where $L$  is the overall degrees of freedom of the system. Similar to the emergence of Fisher zeros in the equilibrium phase transition, DQPTs can occur at some critical times $t^*$, where $\mathcal{L}(t)$ vanishes, and the corresponding dynamical free energy $f(t)$ exhibits divergent behavior in the thermodynamic limit.

To illustrate our quantum quench protocol, cross-stitch networks consisted of two interconnected chains are plotted in Fig.~\ref{001}(a), the unit cell of which is given by two lattice sites $(A,B)$, and the wave function at the $n$-th unit cell is denoted by $\Psi_n$.
The stationary Schr\"{o}dinger equation $\hat H \Psi_n = E\Psi_n$ is expressed as~\cite{Flach1}
\begin{equation}
\hat{\epsilon_n}\Psi_n - \hat{V}\Psi_n - \hat{T}(\Psi_{n-1} + \Psi_{n+1}) = E\Psi_n\ ,
\label{Eq3}
\end{equation}
with
\begin{equation}
\hat{\epsilon_n} = \left(
\begin{array}{ccc}
\epsilon_n^a & 0 \\
0 & \epsilon_n^b
\end{array} \right),\quad
\hat{V} = \left(
\begin{array}{ccc}
0 & V \\
V & 0
\end{array} \right),\quad
\hat{T} = \left(
\begin{array}{ccc}
J & J \\
J & J
\end{array} \right)\;.
\label{Eq4}
\end{equation}
where $J$ is the inter-cell hopping strength and $V$ is the intra-cell hopping strength. In the absence of the potential, $\epsilon_n^a=\epsilon_n^b=0$, there is exactly one flat band $E_{FB} = V$, associated with an antisymmetrical CLS $|f_{m}\rangle=(-1,1)^T \delta_{n,m} / \sqrt{2}$, and one dispersive band $E(k) = -4 J \cos(k) - V$, associated with a Bloch wave function $|k\rangle=e^{i k n}u_k(n)$ with $u_k(n)$ being the periodic envelope function, as shown in Fig.~\ref{002}(a). Here we choose $J=1$ and $V=4$.

To ensure the occurrence of DQPTs we conceive that a symmetrical CLS need to be constructed such that the antisymmetrical CLSs can be exactly eliminated by destructive overlapping.
 We find that when $J=0$ the original cross-stitch lattice can be transformed into an isolated two-site lattice, as shown in Fig.~\ref{001}(b). The bulk momentum-space Hamiltonian in Eq.(\ref{Eq3}) becomes $\hat H(k)=\hat{V}$, independent of the quasimomentum $k$. As shown in Fig.~\ref{002}(b) ($J=0$ and $V=4$), we obtain two flat bands $E_{\pm} = \pm V$ with the corresponding eigenstates $|\phi_\pm\rangle = (\mp1,1)^T \delta_{n,m}/\sqrt{2}$ being antisymmetrical/symmetrical CLSs respectively.

We first consider the quench process with $J$ changing as $J=0\rightarrow J\neq0$, a single particle is initially prepared in the ground state $\ket{\Psi_g}$ of $\hat H(J=0)$.
Without loss of generality, the pre-quench quantum state $\ket{\Psi_g}= |\phi_-\rangle= (1,1)^T \delta_{n,m}/\sqrt{2}$ is localized at the $m$-th unit cell. Performing a sudden quench, the Loschmidt amplitude can be written as
\begin{equation}
\begin{aligned}
  \mathcal{G}(t) &= \langle \Psi_g|e^{-i\hat H(J\neq0) t}|\Psi_g\rangle\\
 & = \sum_{m'} e^{-i V t}|\langle f_{m'}|\Psi_g\rangle|^2+\sum_k e^{i(4J\cos(k)+V) t}|\langle k|\Psi_g\rangle|^2.
 \label{Eq5}
\end{aligned}
\end{equation}
Due to the fact that the CLS $|f_{m'= m}\rangle$ is antisymmetrical while $\ket{\Psi_g}$ is symmetrical, we have the relation $\langle f_{m'= m}|\Psi_g\rangle=0$. For the CLS $|f_{m'\neq m}\rangle$, recalling the existence of a complete orthogonal set of the CLSs on the cross-stitch lattice, then we have $|f_{m'\neq m}\rangle$ are orthogonal to $|\Psi_g\rangle$, i.e., $\langle f_{m'\neq m}|\Psi_g\rangle=0$. So the first term on the right-hand-side of Eq.(\ref{Eq5}) vanishes, i.e., $\sum_{m'} e^{-i V t}|\langle f_{m'}|\Psi_g\rangle|^2=0$.

In general the periodicity of $u_k(n)$ varies with the different quasimomentum $k$. So, only $|\langle k|\Psi_g\rangle|^2$ with minimum $k$ can be approximated as $\frac{1}{L}$, while others can not. However, according to the fact $\overline{|\langle k|\Psi_g\rangle|^2}=\frac{\sum_k |\langle k|\Psi_g\rangle|^2}{L}=\frac{1}{L}$, we approximately obtain
\begin{align}\label{Eq6}
  \mathcal{G}(t) \approx \sum_k e^{i(4J\cos(k)+V) t}\overline{|\langle k|\Psi_g\rangle|^2} =\frac{1}{L}\sum_k e^{i(4J\cos(k)+V) t}.
\end{align}
In the large $L$ limit, since the quasimomentum $k$ continuously distributes within $(0,2\pi)$, we can replace the summation by integration
\begin{align}\label{Eq7}
  \mathcal{G}(t) = \frac{1}{2\pi}\int_{0}^{2\pi}e^{i(4J\cos(k)+V) t} d k = e^{i V t} J_0(4J t),
\end{align}
where $J_0(4J t)$ is the zero-order Bessel function. It is known that $J_0(x)$ has a series of zeros $x_i$ with $i=1, 2, 3, \cdots$, which indicates that the Loschmit echo becomes zero at times
\begin{equation}\label{Eq8}
t_i^{*}=\frac{x_i}{4 J}.
\end{equation}

\begin{figure}
  \centering
  \includegraphics[width=0.5\textwidth]{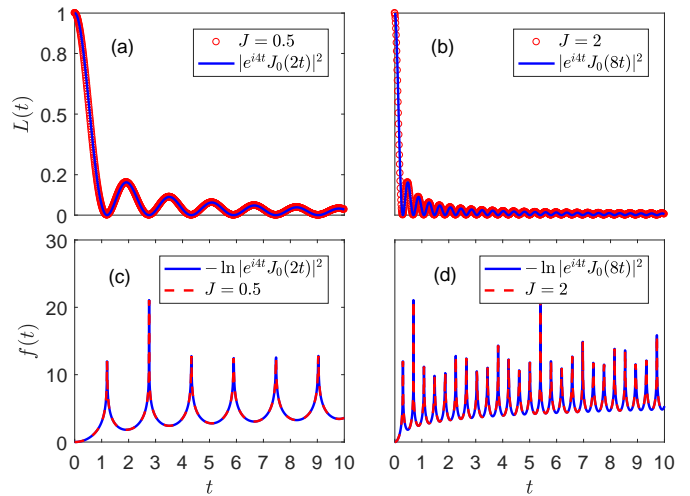}
  \caption{(Color online) The panels (a) and (b) plot the Loschmidt echo $\mathcal{L}(t)$ for different quench parameters $J$.
 At a critical time $t_i^{*}=\frac{x_i}{4 J}$, $\mathcal{L}(t)$ (red hollow circle) reaches the zero point, which agrees with the behaviors of the analytic result $|e^{i V t} J_0(4J t)|^2$ (blue solid line). The panels (c) and (d) plot the dynamical free energy $f(t)$ for different quench parameters $J$.
 At a critical time $t_i^{*}=\frac{x_i}{4 J}$, $f(t)$ (red double dash line) exhibits a sharp peak, which also agrees with the behaviors of the analytic result $-\ln|e^{i V t} J_0(4J t)|^2$ (blue solid line). Here we apply the periodic boundary condition and choose $V=4$. The total number of unit cells is set to be $L = 1000$.}
  \label{003}
\end{figure}

To strengthen the validity of our analytical results, we numerically study the Loschmidt echo and the dynamical free energy.
The initial state is set to be the ground state of $\hat H(J=0)$, and then a finite $J$ is switched on at $t=0$.
According to the theory of DQPTs, the occurrence of a series of zeros in the Loschmidt echo can be recognized as the signatures of DQPTs, and we focus our attention on it first. 
Without loss of generality, we choose a symmetrical CLS $|\phi_-\rangle$ and calculate $\mathcal{L}(t)$ with different $J$'s. As shown in Figs.~\ref{003}(a) and (b), the Loschmidt echo does become zero at some critical times of which the values agree very well with the analytic prediction Eq.(\ref{Eq8}). This demonstrates that Eq.(\ref{Eq6}) is a good approximation.

To show the zeros of $\mathcal{L}(t)$ more reliably, we also calculate the dynamical free energy $f(t)$, which diverges at the dynamical critical time. As shown in Figs.~\ref{003}(c) and (d), the numerical and analytical results are in good agreement with each other, $f(t)$ does exhibit obvious peaks at $t=t_i^{*}$. We also implement calculations for various $J$'s and obtain similar results as expected.
\begin{figure}
  \centering
  \includegraphics[width=0.5\textwidth]{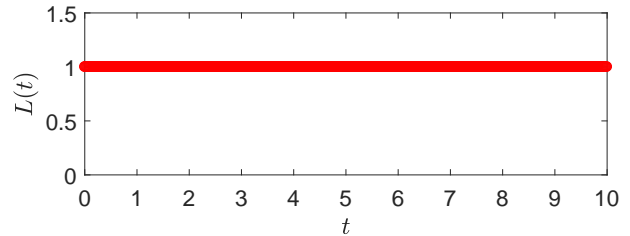}
  \caption{(Color online) The Loschmidt echo $\mathcal{L}(t)$ in the quench process from $\hat H(J=1)$ to $\hat H(J=0)$. Here $V=4$, $L=1000$, and the periodic boundary condition is adapted.}
  \label{004}
\end{figure}

Here we emphasize that (anti)symmetric properties of the CLSs and the existence of a complete orthogonal set of the CLSs are both crucial to ensure $\sum_{m'} e^{-i V t}|\langle f_{m'}|\Psi_g\rangle|^2=0$,
neither one can be absent. 
In addition, the approximate substitution $|\langle k|\Psi_g\rangle|^2 \rightarrow \overline{|\langle k|\Psi_g\rangle|^2}$ requires a simple structure of the real space wave function $|k\rangle$. On the cross-stitch lattice, the amplitudes of $|k\rangle$ on two sites $(A,B)$ are equal, so this substitution is a good approximation. However, in other $U=1$ flat band networks, such as the diamond lattice~\cite{Flach6}, the amplitudes of $|k\rangle$ on three sites of the unit cell are different from each other, this approximation is no longer valid, and DQPTs do not occur as expected.

\section{a non-trivial example of $\mathcal{L}(t)\equiv1$}
\label{n3}
In this section we study the converse quench process with $J\neq0\rightarrow J=0$. No DQPTs are found, however we find an interesting feature in this process. In general the Loschmidt echo $\mathcal{L}(t)$ gradually decreases to zero as the time is long enough, except in the special case that the pre-quench quantum state is an eigenstate of the post-quench Hamiltonian. It can be easily proved that if $\hat H_f \ket{\Psi_i}=E_i \ket{\Psi_i}$, we must have $\mathcal{G}(t)= \bra{\Psi_i}e^{-i\hat H_f t} \ket{\Psi_i}=\sum_n e^{-i E_n t}|\langle \Psi_n|\Psi_i\rangle|^2= e^{-i E_i t}$.
Thus, the Loschmidt echo $\mathcal{L}(t)\equiv1$. Now, here comes a natural question whether it can be deduced from $\mathcal{L}(t)\equiv1$ that the pre-quench quantum state is the eigenstate of the post-quench Hamiltonian?
It seems that the answer is no, however, neither a strict mathematical proof nor an explicit counterexample is reported until now. Here we give a non-trivial example to answer this question. In this process, a single particle is initially prepared in the ground state $|k\rangle=e^{i k n}u_k(n)$. We emphasize that the structure of $|k\rangle$ plays a crucial role in the time evolution, which will be demonstrated later in details. Under the Fourier transformation, the real space Hamiltonian $\hat H(J\neq0)$ can be written in momentum space as
\begin{equation}
\label{Eq9}
H_{k}=\sum_{k}
\vec C_{k}^{\phantom{'}\dag}
\mathcal{H}_{k}
\vec C_{k},
\end{equation}
where  the ``spinor'' $\vec C_{k}=[c_{A,k},c_{B,k}]^{\rm T}$ represents the two sites in the unit cell and
\begin{equation}
\label{Eq10}
\mathcal{H}_{k}=
\begin{bmatrix}
-2J \cos(k) & -V-2J \cos(k)\\
-V-2J \cos(k)& -2J \cos(k)
\end{bmatrix}.
\end{equation}
By diagonalizing the Hamiltonian $\mathcal{H}_{k}$, we get one flat band with the corresponding eigenstate $|\psi_{FB}\rangle=[-1,1]^{\rm T}$ and one dispersive band with the corresponding eigenstate $|\psi_k\rangle=[1,1]^{\rm T}$. This means the amplitudes of the wave function $|\psi_k\rangle$ on two sites $(A,B)$ in the momentum space are equal. By applying the inverse Fourier transformation of $|\psi_k\rangle$, the real space wave function $|k\rangle$ has the form $[\cdots,u_{n-1},u_{n-1},u_n,u_n,,u_{n+1},u_{n+1},\cdots]^{\rm T}, 1\leq n\leq L$.

Now performing a sudden quench, the Loschmidt amplitude can be written as
\begin{equation}
\begin{aligned}
  \mathcal{G}(t) &= \langle k|e^{-i\hat H(J=0) t}|k\rangle\\
 & = \sum_k e^{-i V t}|\langle \phi_+|k\rangle|^2+\sum_k e^{i V t}|\langle \phi_-|k\rangle|^2.
 \label{Eq11}
\end{aligned}
\end{equation}
The overlap between an antisymmetrical CLS $|\phi_+\rangle = (-1,1)^T \delta_{n,m}/\sqrt{2}$ and $|k\rangle$ must vanish, i.e., $|\langle \phi_+|k\rangle|^2=0$, while the squared overlap between a symmetrical CLS $|\phi_-\rangle = (1,1)^T \delta_{n,m}/\sqrt{2}$ and $|k\rangle$ sums to 1, i.e., $\sum_k |\langle \phi_-|k\rangle|^2=1$. Finally we obtain
\begin{align}
  \mathcal{G}(t) = e^{i V t}.
 \label{Eq12}
\end{align}

Thus, our example shows that the pre-quench quantum state is not an eigenstate of the post-quench Hamiltonian, whereas the Loschmidt echo $\mathcal{L}(t)\equiv1$. We also numerically verify our analytic prediction in Fig.~\ref{004}, where the numerical results agree well with the analytic results.

\section{Conclusions}
\label{n4}
In summary, we have studied the quench dynamics on cross-stitch flat band networks by preparing the initial state as an eigenstate of the initial Hamiltonian $H(J_i)$ and then performing a sudden quench to the final Hamiltonian $H(J_f)$. For the quench process changing as $J=0\rightarrow J\neq0$, we calculate the Loschmidt echo both analytically and numerically. We find there exist a series of zero points at critical times $t^{*}$, at which the DQPTs occur. We further study the converse quench process with $J\neq0\rightarrow J=0$, and find that Loschmidt echo $\mathcal{L}(t)\equiv1$ during the whole process and the pre-quench quantum state is not an eigenstate of the post-quench Hamiltonian. We believe that our findings will enrich the studies of DQPTs.

Finally, we would like to point out that this nonequilibrium scenario can be realized in the ultracold-atom experiment. The Lieb lattice~\cite{lieb1,lieb2}, which hosts a variety of novel phenomena when interactions are introduced, has been realized as the prototypical model for exploring flat band in the ultracold-atom system since it is relatively simple to transfer atoms into the flat band.
The cross-stitch lattice has not been realized experimentally yet, which is partly due to the difficulty in transferring atoms into the flat band (the upper energy band). In our quantum quench protocol, the initial quantum state is prepared as the eigenstate of the lower energy band, which is more accessible in the ultracold-atom experiment.
The quench operations can be realized by drastically increasing or decreasing the spacing of unit cells, which leads either $J=0$ or $J\neq0$.
By using time- and momentum-resolved full state tomography methods, the dynamical evolution of the wave function in optical lattices can be monitored, hence the observation of DQPTs on the cross-stitch lattice can be realized experimentally.

\begin{acknowledgments}
This work was supported by the National Natural Science Foundation of China (Grant No. 11674051), the Fundamental Research Funds for the Central Universities, and Postgraduate Research $\&$ Practice Innovation Program of Jiangsu Province (Grant No. KYCX18\_0057).

\end{acknowledgments}



\begin{thebibliography}{36}
\bibitem{Heyl1} M. Heyl, Rep. Prog. Phys. {\bf 81}, 054001 (2018).
\bibitem{Heyl2} M. Heyl, Phys. Rev. Lett. {\bf 113}, 205701 (2014).
\bibitem{Heyl3} M. Heyl, Phys. Rev. Lett. {\bf 115}, 140602 (2015).
\bibitem{FA} F. Andraschko and J. Sirker, Phys. Rev. B {\bf 89}, 125120 (2014).
\bibitem{Eck} E. Canovi, P. Werner, and M. Eckstein, Phys. Rev. Lett. {\bf 113}, 265702 (2014).
\bibitem{CK} C. Karrasch and D. Schuricht, Phys. Rev. B 87, 195104 (2013).
\bibitem{MM} M. Marcuzzi, E. Levi, S. Diehl, J. P. Garrahan, and I. Lesanovsky, Phys. Rev. Lett. {\bf 113}, 210401 (2014).
\bibitem{Jpg} J. M. Hickey, S. Genway, and J. P. Garrahan, Phys. Rev. B {\bf 89}, 054301 (2014).
\bibitem{Lang} J. Lang, B. Frank, and J. C. Halimeh, Phys. Rev. Lett. {\bf 121}, 130603 (2018).

\bibitem{Fisher} M. E. Fisher, in Boulder Lectures in Theoretical Physics(University of Colorado, Boulder, 1965) Vol 7.
\bibitem{Heyl} M. Heyl, A. Polkovnikov, and S. Kehrein, Phys. Rev. Lett. {\bf 110}, 135704 (2013).
\bibitem{Brandner} K. Brandner, V. F. Maisi, J. P. Pekola, J. P. Garrahan, and C. Flindt, Phys. Rev. Lett. {\bf 118}, 180601 (2017).
\bibitem{Marzolino} U. Marzolino and T. Prosen, Phys. Rev. B {\bf 96}, 104402 (2017).
\bibitem{Vajna1} S. Vajna and B. D\'{o}ra, Phys. Rev. B {\bf 89}, 161105 (2014).

\bibitem{Vajna2} S. Vajna and B. D\'{o}ra, Phys. Rev. B {\bf 91}, 155127 (2015).
\bibitem{Heyl4} J. C. Budich and M. Heyl, Phys. Rev. B {\bf 93}, 085416 (2016).
\bibitem{avb} Z. Huang and A. V. Balatsky, Phys. Rev. Lett. {\bf 117}, 086802 (2016).
\bibitem{chen1} C. Yang, L. Li, and S. Chen, Phys. Rev. B {\bf 97}, 060304(R) (2018).
\bibitem{sch} M. Schmitt and S. Kehrein, Phys. Rev. B {\bf 92}, 075114 (2015).
\bibitem{Sedlmayr} N. Sedlmayr, P. Jaeger, M. Maiti, and J. Sirker, Phys. Rev. B {\bf 97}, 064304 (2018).

\bibitem{wang} H. Yin, S. Chen, X. Gao, and P. Wang, Phys. Rev. A {\bf 97}, 033624 (2018).
\bibitem{chen2} C. Yang, Y. Wang, P. Wang, X. Gao, and S. Chen, Phys. Rev. B {\bf 95}, 184201 (2017).

\bibitem{Jurcevic} P. Jurcevic, H. Shen, P. Hauke, C. Maier, T. Brydges, C. Hempel, B. P. Lanyon, M. Heyl, R. Blatt, and C. F. Roos, Phys. Rev. Lett. {\bf 119}, 080501 (2017).
\bibitem{Fl} N. Fl\"{a}schner, D. Vogel, M. Tarnowski, B. S. Rem, D.-S. L\"{u}hmann, M. Heyl, J. C. Budich, L. Mathey, K. Sengstock, and C. Weitenberg, Nat. Phys. {\bf 14}, 265 (2018).


\bibitem{Flach1} S. Flach, D. Leykam, J. D. Bodyfelt, P. Matthies, and A. S. Desyatnikov, Europhys. Lett. {\bf 105}, 30001 (2014).
\bibitem{Flach2} W. Maimaiti, A. Andreanov, H. C. Park, O. Gendelman, and S. Flach, Phys. Rev. B {\bf 95}, 115135 (2017).
\bibitem{Flach3} A. Ramachandran, A. Andreanov, and S. Flach, Phys. Rev. B {\bf 96}, 161104(R) (2017).
\bibitem{Flach4} J. D. Bodyfelt, D. Leykam, C. Danieli, X. Yu, and S. Flach, Phys. Rev. Lett. {\bf 113}, 236403 (2014).
\bibitem{Flach5} R. Khomeriki and S. Flach, Phys. Rev. Lett. {\bf 116}, 245301 (2016).
\bibitem{Gneiting} C. Gneiting, Z. Li, and F. Nori, Phys. Rev. B {\bf 98}, 134203 (2018).

\bibitem{Vidal1} J. Vidal, R. Mosseri, and B. Doucot, Phys. Rev. Lett. {\bf 81}, 5888 (1998).
\bibitem{Vidal2} J. Vidal, B. Doucot, R. Mosseri, and P. Butaud, Phys. Rev. Lett. {\bf 85}, 3906 (2000).

\bibitem{LIU} Z. Liu, E. J. Bergholtz, H. Fan, and A. M. L\"{a}uchli, Phys. Rev. Lett. {\bf 109}, 186805 (2012).
\bibitem{chen} L. Chen, T. Mazaheri, A. Seidel, and X. Tang, J. Phys. A: Math. Theor. {\bf 47}, 152001 (2014).

\bibitem{Flach6} C. Danieli, J. D. Bodyfelt, and S. Flach, Phys. Rev. B {\bf 91}, 235134 (2015).
\bibitem{Flach7} A. R. Kolovsky, A. Ramachandran, and S. Flach, Phys. Rev. B {\bf 97}, 045120 (2018).
\bibitem{Flach8} C. Danieli, A. Maluckov, and S. Flach, J. Low Temp. Phys. {\bf 44}, 678 (2018).
\bibitem{MRN} M. R\"{o}ntgen, C. V. Morfonios, and P. Schmelcher, Phys. Rev. B {\bf 97}, 035161 (2018).
\bibitem{Perchikov} N. Perchikov and O. V. Gendelman, Phys. Rev. E {\bf 96}, 052208 (2017).
\bibitem{Vicencio1} M. Johansson, U. Naether, and R. A. Vicencio, Phys. Rev. E {\bf 92}, 032912 (2015).
\bibitem{Vicencio2} B. Real and R. A. Vicencio, Phys. Rev. A {\bf 98}, 053845 (2018).

\bibitem{Rhim} J.-W. Rhim and B.-J. Yang, Phys. Rev. B {\bf 99}, 045107 (2019).


\bibitem{Huber} S. D. Huber and E. Altman, Phys. Rev. B {\bf 82}, 184502 (2010).
\bibitem{Lieb} E. H. Lieb, Phys. Rev. Lett. {\bf 62}, 1201 (1989).

\bibitem{lieb1} H. Ozawa, S. Taie, T. Ichinose, and Y. Takahashi, Phys. Rev. Lett. {\bf 118}, 175301 (2017).
\bibitem{lieb2} A. Julku, S. Peotta, T. I. Vanhala, D. H. Kim, and P. T\"{o}rm\"{a}, Phys. Rev. Lett. {\bf 117}, 045303 (2016).

\end{thebibliography}
\end{document}